\documentclass[12pt]{article}

\begin{document}

\author{A.I.Volokitin$^{1,2}\footnote{Corresponding author:
Email avoli@samgtu.ru}$ and B.N.J.Persson$^1$ \\
\\
$^1$Institut f\"ur Festk\"orperforschung, Forschungszentrum \\
J\"ulich, D-52425, Germany\\
$^2$Samara State Technical University, 443100 Samara,\\
Russia}
\title{Adsorbate vibrational mode enhancement of radiative heat 
transfer and van der Waals friction}
\maketitle

\begin{abstract}
We study the dependence of the heat transfer and the van der Waals friction 
between two semi-infinite
solids on the dielectric properties of the bodies. We show that the heat
transfer and  van der Waals friction at short separation between 
the solids may increase by many orders
of magnitude when the surfaces are covered by adsorbates, or can support
low-frequency surface plasmons. In this case the heat transfer and 
van der Waals friction are determined
by resonant photon tunneling between adsorbate vibrational modes, or surface
plasmon modes. The enhancement of the van der Waals friction is especially 
large when in the adsorbed layer 
there is an acoustic branch for the vibrations parallel to the surface 
like in the case of Cs adsorption on Cu(100) surface. In this case we show 
that even for  separation $d=10$nm, the  van der Waals friction induced by 
adsorbates can be so large that it can be measured with the present 
state-of-art equipment.
The van an der Waals friction is characterized by a strong distance 
dependence ($\sim 1/d^6$), and at the small distances it can be much larger than 
\textit{the electrostatic} friction observed in \cite{Stipe}.  
\end{abstract}

\section{ Introduction}

In the history of physics the studies of thermal radiation from  materials 
always played a very important role. Here, it is enough to mention that quantum 
mechanics 
originated from the attempts  to  explain a paradoxical experimental results related to the  black body radiation. 
In the past, the nonradiative near-field part of electromagnetic radiation 
usually was 
 ignored, because it plays no role in the far-field properties of emission from planar 
sources. Nevertheless, recent interest in microscale and nanoscale radiative 
heat transfer 
\cite{Van Hove,Levin1,Pendry,Volokitin1,Volokitin2}, together 
with the development of local-probe thermal microscopy 
\cite{Majumdar} have raised new challenges. These topics, and the  
progress in detecting  non-contact friction  on sub-attonewton  level 
\cite{Dorofeev,Gostmann,Stipe,Mamin,Hoffmann}, and the observation of 
coherent thermal emission from doped silicon and 
silicon carbide (SiC) gratings \cite{Greffet}  have in common the substantial 
role of the nonradiative (evanescent) 
thermal field.

It is well known that the
radiative heat transfer between two bodies separated by $d>>\lambda_T=c\hbar /k_BT$
 is given  by the Stefan-Boltzmann
law: 
\begin{equation}
S=\frac{\pi ^2k_B^4}{60\hbar ^3c^2}\left( T_1^4-T_2^4\right) ,
\label{stefan}
\end{equation}
where $T_1$ and $T_2$ are the temperatures of solid $1$ and $2$,
respectively. In this limiting case the heat transfer between two 
black bodies is 
determined by the propagating electromagnetic waves  radiated by the bodies, 
and does not depend
on the separation $d$.
For $d<\lambda_T$ the heat transfer increases by many orders
of magnitude due to the evanescent electromagnetic waves that decay
exponentially into the vacuum; this is often refereed to as photon
tunneling. At low temperatures  (a few K) it is possible for this 
 form of heat transfer  to be dominant even at spacing of a few mm, see table 1.

\vskip 0.3cm
\textbf{Table 1}. Critical distance $\lambda_T$ as a function of temperature. 
For surface separation $d<\lambda_T$ the heat transfer is dominated by the contribution from 
the evanescent electromagnetic modes. At distances of a few nanometers, radiative heat flow is almost entirely due to evanescent modes.

\begin{tabular}{lr} \hline\hline
$T$(K) & $\lambda_T$($\mu$m) \\
\hline
1 & 2298.8 \\
4.2 & 545.2 \\
100 & 22.9 \\
273 & 8.4 \\
1000 & 2.3 \\
\hline\hline
\end{tabular}

\vskip 0.3cm
The problem of the radiative heat transfer between two flat surfaces was 
considered some years ago by Polder and Van Hove \cite{Van Hove}, Levin and
Rytov \cite{Levin1} and more recently by Pendry \cite{Pendry}, and by Volokitin
and Persson \cite{Volokitin1,Volokitin2}.
 In \cite{Pendry,Volokitin1,Volokitin2}
it was shown that the heat flux can be greatly enhanced if the electrical 
conductivities of
the materials are chosen such as to maximize the heat flow due to photon tunneling. At
room temperature the heat flow is maximal for materials with 
 conductivities corresponding to
semi-metals. In fact, only a thin film ($\sim 10${\AA }) of a
high-resistivity material is needed to maximize the heat flux \cite{Volokitin1}. Another
enhancement mechanism of the radiative heat transfer may arise from
resonant photon tunneling between  localized  states on the different surfaces, 
if the frequency of these
modes is sufficiently low to be excited by thermal radiation. 
Recently, it was predicted enhancement of the heat transfer due to resonant 
photon tunneling
between surface plasmon modes localized on the surfaces of the
semiconductors \cite{Mulet,Volokitin2} and adsorbate vibrational modes 
 \cite{Volokitin2}. 

The problem of radiative heat transfer is closely related to the 
non-contact friction between
nanostructures, including, for example, the frictional drag force between
two-dimensional quantum wells \cite{Gramila1,Gramila2,Sivan} , and the
friction force between an atomic force microscope tip and a substrate \cite
{Dorofeev,Gostmann,Stipe,Mamin,Hoffmann}

Recently several groups have observed  unexpectedly large long-range non-contact 
friction
\cite{Dorofeev,Gostmann,Stipe,Mamin,Hoffmann}.   
 The friction force $F$ acting on an atomic force microscope tip was found to be 
proportional to the velocity $v$, $F=\Gamma v$. At the separation 
$d=100${\AA}  the friction coefficient $\Gamma \approx 10^{-10}-
10^{-13}$kg/s. Although the non-contact friction must have an electromagnetic origin,
 the 
detailed mechanisms is not fully understood yet.   

In a recent Letter, Dorofeyev \textit{et.al.} \cite{Dorofeev} claim that 
the non-contact friction  observed in \cite{Dorofeev,Gostmann} is due to
Ohmic losses mediated by the fluctuating electromagnetic field . This result
is controversial, however, since the van der Waals friction has
been shown \cite{Volokitin3,Persson and Volokitin} to
be many orders of magnitude smaller than the friction  observed by Dorofeyev \textit{%
et.al.} However,   we have shown that when the surfaces are separated by 
about 1nm, the van der
Waals friction can be greatly enhanced by resonant photon
tunneling \cite{Volokitin4,Volokitin5}. Thus, we found that resonant photon 
tunneling between
two semiconductors surfaces of SiC, which can support surface  plasmon modes
in mid-infrared region, may give rise a 1000-fold (or more) enhancement of 
the  friction,
in comparison with the case of good conductors. Furthermore, resonant photon tunneling
between two Cu(100) surface covered by 0.1 monolayer of K may result in  six
 orders of magnitude enhancement of friction compared to clean surfaces.
 At the separation
1nm we obtained a friction comparable in magnitude with the friction observed 
in the experiment
 \cite{Stipe}.

The origin of the van der Waals friction is very closely connected with 
ordinary (attractive) van
der Waals interaction. The van der Waals interaction between atoms (or molecules) 
arises from quantum fluctuations in the electric dipole moment of atoms. 
The short-lived atomic polarity can induce a dipole moment in a
neighboring atom or molecule at some distance away. The same is true  for extended
solids, where thermal and quantum fluctuation of the current density in one
body induces a current density in another body. 
 When two bodies are in relative
motion, the induced current will lag slightly behind the  fluctuating current 
inducing it,
this lag is the origin of the van der Waals friction. 

In contrast to the van der Waals interaction, for which a well established 
theory exist \cite
{Lifshitz}, the 
field of van der Waals friction is still controversial. Thus different authors 
\cite{Theodorovich,Levitov,Polevoi,Mkrtchian,Dedkov,Dedkov1} 
have derived expression for the  van der Waals friction between two flat surfaces and 
between a small particle and flat surface using 
different methods, and obtained results,
which were not confirmed in subsequent calculations \cite{Schaich,Pendry1,
Persson and Zhang,Volokitin3,Volokitin6}.

In \cite{Volokitin3} we have developed the theory of the van der Waals friction 
based on  a dynamical modification of the well known Lifshitz theory \cite{Lifshitz1} 
of the van der Waals interaction. In the nonretarded limit, and for zero temperature, 
this theory agree with the results of Pendry \cite{Pendry1}.  We have also calculated 
the van der Waals friction between two flat surfaces in normal relative motion
 \cite{Volokitin4,Volokitin5},  and 
found a drastic difference in comparison with parallel relative motion. In the 
limit, when one of the bodies is sufficiently rarefied,  our theory  
gives the  friction between flat surface and small particle, which agrees with the 
results of Tomassone and Widom \cite{Tomassone}.             

The organization of this article is as follows. In Sec.2 we present a short overview 
of the main principles of the radiative heat transfer, focusing mainly on the  adsorbate 
vibrational mode enhancement of the radiative heat transfer. In Sec.3 we study  
the van der Waals friction, and its enhancement due to surface polaritons and 
adsorbates. In the light of our 
theoretical results, we also  
   discuss  
non-contact friction experiments. Finally,  Sec.4 contains the summary.

\section{Radiative heat transfer}
\subsection{Clean surfaces}

According to \cite{Van Hove,Levin1,Pendry,Volokitin1} the heat transfer
between two semi-infinite bodies, separated by a vacuum gap with the width $%
d$, is given by the formula 
\begin{equation}
S=\int_0^\infty d\omega \left( \Pi _1-\Pi _2\right) M
\end{equation}
where 
\[
M=\frac 1{4\pi ^2}\int_0^{\omega /c}dq\,q\frac{(1-\mid R_{1p}(\omega )\mid
^2)(1-\mid R_{2p}(\omega )\mid ^2)}{\mid 1-\mathrm{e}^{2ipd}R_{1p}(\omega
)R_{2p}(\omega )\mid ^2}
\]
\begin{equation}
+\frac 1{\pi ^2}\int_{\omega /c}^\infty dq\,q\mathrm{e}^{-2kd}\times \frac{%
\mathrm{Im}R_{1p}(\omega )\mathrm{Im}R_{2p}(\omega )}{\mid 1-\mathrm{e}%
^{-2\mid p\mid d}R_{1p}(\omega )R_{2p}(\omega )\mid ^2}+\left[ p\rightarrow
s\right]   \label{heat},
\end{equation}
and where the symbol $\left[ p\rightarrow s\right] $ stands for the terms
which  are obtained from the first two terms by replacing the reflection
coefficient $R_p$ for  $p$-polarized electromagnetic waves with the reflection
coefficient $R_s$ for $s$- polarized electromagnetic waves, and where $%
p=((\omega /c)^2-q^2)^{1/2}$, and $ k=|p|$. The Planck function of solid \textbf{1}
\begin{equation}
\Pi _1(\omega )=\hbar \omega \left( e^{\hbar \omega /k_BT_1}-1\right) ^{-1},
\end{equation}
and similar for $\Pi _2$. The contributions to the heat transfer from the
propagating ($q<\omega /c$) and evanescent ($q>\omega /c$) electromagnetic
waves are determined by the first and the second terms in Eq.(\ref{heat}),
respectively.  

From Eq.(\ref{heat}) it is easy to see that 
  the propagating photon modes give the main contribution to the heat transfer for
$q<\lambda_T^{-1}$,  
while  the evanescent modes  for $q<1/d$. 
 Thus from phase space 
arguments it follows that the number of the channels of the heat transfer available 
for evanescent waves will be by a factor $(\lambda_T/d)^2$ larger than the 
number of the channel  available for propagating waves. 
At $d=1$nm and $T=300$K, this ratio 
is of the order $\sim 10^8$.   

Let us firstly consider some general consequences of Eq. (\ref{heat}). In the
case of the heat transfer through free photons ($q \le \omega/c$), the transfer
is maximal when both the bodies are perfectly black and have zero reflection
amplitude $R=R_r+iR_i=0$. 
The heat flux due to the evanescent waves is a maximal when \cite{Pendry} 
\begin{equation}
R_r^2+R_i^2=\mathrm{e}^{2kd}
\end{equation}
 Substituting this result into (\ref{heat}) gives the maximal
contribution from the evanescent waves 
\begin{equation}
(S_z)_{max}^{evan}=\frac{k_B^2T^2q_c^2}{24\hbar}  \label{heleven}
\end{equation}
where $q_c$ is a cut-off in $q$, determined by the properties of the
material. It is clear that the largest possible $q_c \sim 1/b$, where $b$ is
an inter-atomic distance. Thus, from Eqs.(\ref{heleven}) and (\ref{stefan}) 
we get the ratio of the maximum heat flux connected with evanescent waves to 
the heat flux due to black body radiation $(S_z)_{max}/S_{BB}\approx 0.25\cdot
(\lambda_T/b)^2$. Thus, at room temperature the contribution 
to the heat flux from evanescent waves may be   eight orders of magnitude larger than 
the contribution from the black body radiation, and  
the upper
boundary for the  heat transfer at room temperature: $(S_z)_{max}
\sim 10^{11}\mathrm{Wm^{-2}}$.

Let us now apply the general theory to a few different  materials. For good conductors, 
using Fresnel formulas for reflection coefficient 
and assuming  $k_BT/4\pi \hbar\sigma \ll 1$, and 
$\lambda_T(k_BT/4\pi
\hbar\sigma)^{3/2}
 < d < \lambda_T(k_BT/4\pi
\hbar\sigma)^{- 1/2}$ ($\sigma$ is the conductivity) 
the contribution to the heat transfer 
from $p$-polarized waves for good  
conductors ($k_BT/4\pi \hbar \sigma \ll  1$) is 
given by
\begin{equation}
S_p \approx 0.2\frac{(k_BT)^2}{\hbar\lambda_T d}
\left(\frac{k_BT}{4\pi\hbar \sigma}\right)^{1/2}, 
\label{ppolar}
\end{equation}
and for the metal with lower conductivity ( $k_BT/4\pi \hbar\sigma\le1$), 
for $d<\lambda_T(k_BT/4\pi
\hbar\sigma)^{-1/2}$ we get
\begin{equation} 
S_p \approx 0.12\frac{(k_BT)^2}{\hbar d^2}\left(\frac{k_BT}{4\pi\hbar \sigma}\right)^2
\left(1+\mathrm{ln}\frac{4\pi\hbar \sigma}{k_BT}\right).
\label{1ppolar}
\end{equation}
 For good conductors the 
heat flux depends on the separation  as $\sim d^{-1}$, and increases with decreasing 
 conductivity as $\sigma^{-1/2}$.  
For  $k_BT/4\pi \hbar\sigma\le1$ 
the heat flux decreases with separation as $d^{-2}$, and 
increases with decreasing conductivity as $\sigma^{-2}$, while the 
$s$-wave contribution is distance independent,
$S_s\approx 0.25 k_BT\sigma/\lambda_T^2$.
Fig. 1a shows the heat transfer between two semi-infinite silver bodies
separated
by the distance $d$,  at the temperatures
$T_1 = 273 {\rm K}$ and $T_2 = 0 {\rm K}$. The $s$- and $p$-wave contributions
are shown separately.
The $p$-wave contribution has been calculated using non-local optics, i.e. 
including  
spatial dispersion of the dielectric function (the
dashed
line shows the result using local optics). It is remarkable how important the
$s$-contribution is even for short distances.
 Note from Fig.1a that the local optics
contribution to
$S_p$ depends nearly linearly on $1/d$ in the studied distance interval,
and that this
contribution is much smaller than the $s$-wave contribution. Both these
observations
 agree with  analytical formulas given above.
 However, for the very high-resistivity
materials, the $p$-wave contribution becomes much more important,
and a crossover to a $1/d^2$-dependence of $S_p$ is observed at  short
separations
$d$. This is illustrated in Fig.1b and 1c, which show results for  the
same
parameters as in Fig. 1a, except that the electron mean free path has been
reduced from
$l= 560 \ {\rm \AA}$ (the electron mean free path for silver at room
temperature)
to $20 \ {\rm \AA}$ (roughly the  mean free path in lead at room
temperature)
(Fig. 1b) and $3.4 \ {\rm \AA}$
(of order the lattice constant, representing the minimal possible mean free
path)
(Fig. 1c). Note that when $l$ decreases, the $p$-wave contribution to the heat
transfer increases while
the $s$-wave contribution decreases. Since the mean free path cannot be much smaller
than the lattice constant, the result in Fig. 1c
represent the largest possible $p$-wave
contribution for normal metals. However, the $p$-wave contribution may be even
larger for
other materials, e.g., semimetals, with lower carrier concentration than in
normal metals.
For high resistivity
materials,  when $k_BT/4\pi\hbar\sigma>1$ the heat flux is proportional to the 
conductivity
\begin{equation}
S_p \approx 0.2\frac{k_BT\sigma}{d^2}
\end{equation}
Figure 2 show the thermal flux as a function of the conductivity of the solids. 
Again we have assumed that one body is at zero temperature and the other 
at $T=273K$. The  surfaces are separated by $d=10\ \mathrm{\AA }$.
 The heat flux for other separations can be obtained
using scaling $\sim 1/d^2$ which holds for high-resistivity materials. The heat 
flux 
is  maximal at $\sigma=1316 (\Omega\cdot$ m$)^{-1}$.
Finally, we note that thin high-resistivity coatings can drastically
increase the heat transfer between two solids. This is illustrated in Fig.3, which 
shows the heat flux for the case  when thin films ($\sim 10\,\mathrm{\AA }$) of high
resistivity material, $\rho =0.14\ \Omega\cdot \mathrm{cm}$, are deposited on
silver.  (a) and
(b) shows the $p$- and $s$-contributions, respectively. Also shown are the
heat flux when the two bodies are made from silver, and from the high
resistivity material. It is interesting to note that while the $p$%
-wave contribution to the heat flux for the coated surfaces is strongly
influenced by the coating, the $s$-contribution is nearly unaffected.

Another case where the   heat transfer can be large is when  
 resonant photon tunneling occurs between surface states localized on the 
 different surfaces.     
The resonant condition corresponds to the case when the denominators in Eq.(\ref{heat}) 
are small. For two identical surfaces and $R_i<<1\le R_r$, where $R_i$ and 
$R_r$ are the imaginary and real parts of the reflection coefficient, this 
corresponds to the resonant condition 
\begin{equation}
R_r\exp(-qd)=\pm 1
\label{resonance}
\end{equation}
 This condition  can be fulfilled  in spite of the factor 
 $\exp(-2qd)<1$ because for evanescent 
electromagnetic waves there is no restriction on the magnitude 
of real part or the modulus of $R$. 
This opens up the possibility of resonant denominators  for $R_r^2\gg 1$.  
 Close to the resonance  we can use the approximation 
\begin{equation}
R=\frac {\omega_a}{\omega -\omega _0-i\eta },  \label{six}
\end{equation}
where $\omega_a$ is a constant. Then from the resonant condition 
($R_r=\pm e^{qd}$) we get 
the positions of the resonance 
\[
\omega _{\pm }=\omega _0\pm \omega_ae^{-qd}. 
\]
For the resonance condition  to be valid the separation  $\Delta \omega
=|\omega _{+}-\omega _{-}|$ between two resonances  
must be greater than the width $\eta $ of the
resonance. From this condition we get that the two poles approximation is 
valid for $q\le q_c \approx \ln (2\omega_a/\eta )/d$. 

For $\omega_0>\omega_a$ and $q_cd>1$,  we get 
\begin{equation}
S_{\pm}=\frac{\eta q_c^2}{4\pi}\left[\Pi_1(\omega_0)-\Pi_2(\omega_0)\right].
\label{seven}
\end{equation}

Note, that the explicit $d$ dependence has dropped out of Eq. (\ref{seven}%
). However, $S$ may still be $d$- dependent, through the $d$- dependence of $%
q_c$. For  small distances one can expect that $q_c$ is determined by the
dielectric properties of the material, and thus does not depend on $d$. In
this case the heat transfer will also be  distance independent.

Resonant photon tunneling enhancement of the heat transfer is possible for
two semiconductor surfaces which can support low-frequency surface plasmon
modes in the mid-infrared frequency region. The reflection factor $R_p$ for 
clean semiconductor surface at $d<\lambda_T\left|\epsilon\right|^{-1/2}$ is given
by Fresnel's formula 
\begin{equation}
R_p=\frac{\epsilon-1 }{\epsilon+1 },  \label{eight}
\end{equation}
 where $\epsilon $ is the bulk dielectric function. As an example, 
consider two clean surfaces of silicon carbide (SiC). The optical properties
of this material can be described using an oscillator model \cite{Palik} 
\begin{equation}
\epsilon (\omega )=\epsilon _\infty \left( 1+\frac{\omega _L^2-\omega _T^2}{%
\omega _T^2-\omega ^2-i\Gamma \omega }\right)  \label{nine}
\end{equation}
with $\epsilon _\infty =6.7,\,\omega _L=1.83\cdot 10^{14}s^{-1},\,\omega
_T=1.49\cdot 10^{14}s^{-1},\,$ and $\Gamma =8.9\cdot 10^{11}s^{-1}.$ The
frequency of surface plasmons is determined by condition $\epsilon _r(\omega
_p)=-1$ and from (\ref{nine}) we get $\omega _p=1.78\cdot 10^{14}s^{-1}$. 
The resonance parameters 
\[
\omega_a=\frac{\omega^2_L-\omega_T^2}{\epsilon_\infty \omega_L} =8.2\cdot 10^{12}
\mbox{s}^{-1},\,\eta=\Gamma/2,\,q_c=3.6/d,
\quad\mbox{and}\quad \omega_0\approx\omega_p
\]

 Using these parameters in Eq.(\ref{seven}) and assuming that 
 one surface is at temperature $T=300$ K 
and the other at $0$ K we get the heat flux $S(d)$ between two clean surfaces of 
silicon carbide (SiC): 
\begin{equation}
S_p\approx 8.4\cdot 10^9\frac{1}{d^2} \mbox{W$\cdot$ m$^{-2}$}
\end{equation}
where the distance $d$ is in {\AA}ngstr{\o}m. Note that the heat flux between the two
semiconductor surfaces is several order of magnitude larger than between two
clean good conductor surfaces (see Fig.1).

\subsubsection{Adsorbate vibrational mode enhancement of the radiative heat transfer}

Another  resonant photon tunneling enhancement of the heat transfer  is 
 possible between adsorbate vibrational modes localized on different surfaces. 
Let us consider two particles or adsorbates with  the 
dipole polarizabilities $\alpha_1(\omega)$ and
$\alpha_2(\omega)$ and with the fluctuating dipole moments $p_1^f$ and $p_2^f$  normal
to the surfaces. Accordingly
to the fluctuation-dissipation theorem \cite{Landau},  the  power spectral density
 for the fluctuating dipole
moment is given by
\begin{equation}
\langle p_i^fp_j^f\rangle _{\omega}=\frac{\hbar}{\pi}\left(\frac{1}{2}+n_i(\omega)\right)\mathrm{Im}
\alpha_i(\omega)\delta_{ij}
\label{one}
\end{equation}
where the Bose-Einstein factor
\begin{equation}
n_i(\omega)=\frac{1}{e^{\hbar \omega/k_BT_i}-1}.
\end{equation}
Assume that the particles are situated opposite to each other on two different surfaces, at the  temperatures
$T_1$ and $T_2$, respectively, and separated by the
distance $d$. The fluctuating electric field of  a particle $\mathbf{1}$ does work on a particle $\mathbf{2}$.
The rate of working  is determined
by
\begin{equation}
P_{12}=2\int_0^{\infty}d\omega\,\omega \mathrm{Im}\alpha_2(\omega)\langle E_{12}E_{12}\rangle  _{\omega}
\label{two}
\end{equation}
where  $E_{12}$ is the electric field created by a particle $\mathbf{1}$ at the position of a particle $\mathbf{2}$:
\begin{equation}
E_{12}=\frac{8p_1^f/d^3}{1-\alpha_1 \alpha_2 (8/d^3)^2}
\label{three}
\end{equation}
From Eqs. (\ref{one}-\ref{two}) we get $P_{12}$, and  the rate of cooling of  a particle $\mathbf{2}$
can be obtained
using the same formula by reciprocity. Thus the total heat exchange power  between the particles is given by
\begin{equation}
P=P_{12}- P_{21}=\frac{2\hbar}{\pi}\int_0^{\infty}d\omega\,\omega \frac{\mathrm{Im}\alpha_1\mathrm{Im}\alpha_2(8/d^3)^2}
{\left|1-(8/d^3)^2\alpha_1\alpha_2\right|^2} (n_1(\omega)-n_2(\omega))
\label{four}
\end{equation}

Let us firstly consider some general  consequences of Eq. (\ref{four}).
There are no constrain on the particle polarizability $\alpha(\omega)=\alpha^{\prime}+i\alpha^{\prime \prime}$
 other than that $\alpha^{\prime \prime}$ is positive, and $\alpha^{\prime}$ and $\alpha^{\prime \prime}$ are connected by the
Kramers-Kronig
relation. Therefore, assuming identical surfaces, we are free to maximize the photon-tunneling
transmission coefficient
\begin{equation}
t=\frac{(8\alpha^{\prime \prime}/d^3)^2}{\left|1-(8\alpha /d^3)^2\right|^2}
\label{five}
\end{equation}
This function has a maximum when
\begin{equation}
\alpha^{\prime 2} + \alpha^{\prime \prime 2} =(d^3/8)^2
\end{equation}
so that $t=1/4$. Substituting this result in (\ref{four}) gives the upper bound for
the heat transfer power  between the two particles or adsorbates 
\begin{equation}
P_{max}=\frac{\pi k_B^2}{3\hbar}(T_1^2-T_2^2)
\label{six}
\end{equation}
 For adsorbed molecules at the  concentration $n_a=10^{19}$m$^{-2}$, when one surface is at zero temperatures
and the other is at the room temperature, the maximal heat flux due to the adsorbates $S_{max}=n_aP_{max}=10^{12}$Wm$^{-2}$,
which is nearly 10 order of magnitude larger than the  heat flux due to the black body radiation,
 $S_{BB}=\sigma_B T =4\cdot 10^2$Wm$^{-2}$, where $\sigma_B$ is the Boltzmann 
constant.
The conditions for resonant photon tunneling are determined by equation
\begin{equation}
\alpha^{\prime}(\omega_{\pm})=\pm d^3/8
\label{eight}
\end{equation}
Close to resonance we can  
 use the approximation
\begin{equation}
\alpha\approx\frac{c}{\omega-\omega_0-i\eta},
\label{eleven}
\end{equation}
where $c=e^{*2}/2M\omega_0$, and where $e^*$ and $M$ are the dynamical charge 
and mass of the adsorbate, respectively. Then from the resonant condition (\ref{eight}) we get
\[
\omega_{\pm}=\omega_0\pm 8c/d^3.
\]
The separation between the resonances,   $\Delta\omega =
|\omega_+-\omega_-|$ must be greater than the width $\eta$ of the resonance,
 so that
$8c/d^3>\eta$.

For $\eta<<8c/d^3$,  from Eq. (\ref{four}) we get
\begin{equation}
P=\frac{\hbar \eta}{2} [\omega_+ (n_1(\omega_+)-n_2(\omega_+)) + (+\rightarrow -)].
\label{twelve},
\end{equation}
Using Eq. (\ref{twelve}) we can estimate the heat flux between identical surfaces covered by  adsorbates with concentration $n_a$:
 $S\approx n_a P$.
Interestingly, the  explicit $d$ dependence has dropped out of Eq. (\ref{twelve}).
However, $P$ may still be $d$- dependent, through the $d$- dependence
of $\omega_{\pm}$. For  $\hbar\omega_{\pm}\le k_BT$
 the heat transfer  will be only  weakly distance dependent.
For $8c/d^3<\eta$ we can neglect  multiple scattering of the photons between 
the particles,
so that   the denominator in the integrand in Eq. (\ref{four}) can be approximated 
with unity. For $d>>b$, where
$b$ is the interparticle spacing, the heat flux between two surfaces covered by adsorbates
with concentration $n_{a1}$ and $n_{a2}$ can be obtained after integration of 
the heat flux between
two separated particles. We get
\begin{equation}
S=\frac{24\hbar n_{a1}n_{a2}}{d^4}\int_0^{\infty}d\omega\,\omega \mathrm{Im}\alpha_1 \mathrm{Im}\alpha_2
[n_1(\omega)-n_2(\omega)]
\label{thirteen}
\end{equation}
Assuming that $\alpha$ can be approximated by Eq. (\ref{eleven}),   for $\omega_0<<\eta$  Eq. (\ref{thirteen})
gives  the heat flux between two identical surfaces:
\begin{equation}
S=\frac{12\pi \hbar \omega_0 c^2n_a^2}{d^4\eta}[n_1(\omega_0)-n_2(\omega_0)]
\label{fourteen}
\end{equation}

For the K/Cu(001) system $\omega_0=1.9\cdot10^{13}$s$^{-1}$, and at low coverage 
$e^{*}=0.88e$ \cite{Senet},
which gives  $c=e^{*2}/2M\omega_0=7\cdot10^{-17}$
m$^3$s$^{-1}$. For $\eta = 10^{12}$s$^{-1}$, when one surface has $T=300$K and 
the other  $T=0$K, for $d>b$ and $8a/d^3<\eta$ we get
\begin{equation}
S\approx 5.6\cdot10^{-24}\frac{n_a^2}{d^4}\,\mbox{W$\cdot$ m$^{-2}$}
\end{equation}
where the $d$ is in {\AA}ngstr{\o}m. 

We note that Eq.(\ref{fourteen}) can be obtained directly from the heat
flux between two semi-infinite solids determined by Eq.(\ref{heat}), since
in the limit $d>b$ we can use a macroscopic approach, where all the information about
the optical properties of the surface is included in reflection coefficient \cite
{Volokitin2}.
The reflection  coefficient $R_{p}$, which
take into account the contribution from an adsorbate layer 
is given by \cite{Volokitin}:
\begin{equation}
R_{p}=\frac {1-s/q\epsilon+4\pi n_a[s\alpha_{\parallel}/\epsilon+q \alpha_{\perp}]
-qa(1-4\pi n_aq\alpha_{\parallel})}
{1+s/q\epsilon+4\pi n_a[s\alpha_{\parallel}/\epsilon-q \alpha_{\perp}]
+qa(1+4\pi n_aq\alpha_{\parallel})}, \label{reftwo}
\end{equation}
where
$s=\sqrt{q^2-(\omega/c)^2\epsilon}$,
and  $\alpha_{\parallel}$ and $\alpha_{\perp}$ are the
polarizabilities of adsorbates
in a direction parallel and normal to the surface,  respectively. $\epsilon=1+4\pi i
\sigma/\omega$
is the bulk  dielectric function, where $\sigma$ is a conductivity, and $n_a$ is
the concentration of adsorbates. 
 Eq.(\ref{reftwo}) takes into account that the centers of the adsorbates 
are located at distance $a$ away from image plane of the metal.  
Although  this gives corrections of the order $qa\ll 1$ to the reflection 
 amplitude, for parallel 
 adsorbate vibrations on  the good 
conductors (when $\epsilon\gg 1$), in some cases they  give the most important 
contribution 
to the energy dissipation.  As illustration of this macroscoscopic approach, 
in Fig.(4) we show the $p$-wave contribution to the heat flux for the 
two Cu(100) surfaces covered by a low 
concentration of potassium atoms ($n_a=10^{18}$m$^{-2}$) and the two clean Cu(100) 
surfaces. At separation $d=1$nm the heat flux between two surfaces covered by 
adsorbates is enhanced by five  and three orders of magnitude in comparison with 
the $p$- and $s$-
wave contributions to the heat 
flux between clean surfaces, respectively, and by seven orders of magnitude in 
comparison with the blackbody radiation.

For $d<b$ the macroscopic approach is not valid any more and we must sum the heat 
flux between each pair of the adatoms.
For $\eta = 10^{12}$s$^{-1}$ and $d<10${\AA}, when one surface has
$T=300$K and the other $T=0$K, from Eq.(\ref{twelve})  we get the distance independent
$P\approx10^{-9}$W
. In this case,  for  $n_a=10^{18}$m$^{-3}$
 the heat flux $S\approx Pn_a\approx 10^{9}$Wm$^{-2}$. Under the same conditions 
the $s$-wave contribution to the  heat flux between two clean surfaces
$S_{clean}\approx10^{6}$Wm$^{-2}$. Thus the photon tunneling between the adsorbate vibrational states can
strongly enhance the radiative heat transfer between the surfaces.
However this enhancement of the heat flux disappears if only one of the
surfaces is  covered by adsorbates.

It is interesting to note that in the strong coupling case ($8c/d^3\gg \eta$) the 
heat flux between two molecules does not depend on the dynamical dipole moments 
of the molecules (see Eq.(\ref{twelve}). However, in the opposite case of the 
weak coupling ($8c/d^3\ll \eta$) the heat flux is proportional to the product 
of the squares of the dynamical dipole moments (see Eq.(\ref{fourteen})).

\section{Van der Waals friction}

\subsection{Clean surfaces}

The frictional stress $\sigma_{\perp(\|)}$ which act on the surfaces 
of  two bodies in normal (parallel) relative motion can, to linear order in sliding 
velocity $v$,  be written in the form: $\sigma_{\perp(\|)}=\gamma_{\perp(\|)}v$.
For bodies in parallel relative motion at separation $d<<\lambda_T$ the friction 
coefficient $\gamma_{\|}$ is given by \cite{Volokitin3}

\[
\gamma_{\parallel}=\frac{\hbar}{2\pi^2}\int_0^{\infty}d\omega
\left(-\frac{\partial n}{\partial \omega}\right)\int_{\omega/c}^{\infty}dq\,
q^3e^{-2kd}
\]
\begin{equation}
\times \mathrm{Im}R_{1p}
\mathrm{Im}R_{2p}
\frac1{\left|1-e^{-2kd}R_{1p}R_{2p}\right|^2}
+ [p\rightarrow s].
\label{vdwpar}
\end{equation}
When the two bodies move toward or away from each other the friction 
coefficient is given by \cite{Volokitin4,Volokitin5}
\[
\gamma_{\perp}=\frac{\hbar}{\pi^2}\int_{0}^{\infty}d\omega
\left(-\frac{\partial n}{\partial \omega}\right)\int_{\omega/c}^{\infty}dq\,
qk^2e^{-2kd}
\]
\[
\times \big[\big(\mathrm{Im}R_{1p}+
e^{-2kd}\left|R_{1p}\right|^2\mathrm{Im}R_{2p}\big)\big(\mathrm{Im}R_{2p}+
e^{-2kd}\left|R_{2p}\right|^2\mathrm{Im}R_{1p}\big)
\]
\begin{equation}
+e^{-2kd}\big(\mathrm{Im}\big(R_{1p}R_{2p}\big)\big)^2\big]
\frac 1{\left|1-e^{-2kd}R_{1p}R_{2p}\right|^4}
+[p\rightarrow s],
\label{vdwnor}
\end{equation}
 where the Bose-Einstein factor
\[
n(\omega )=\frac 1{e^{\hbar \omega /k_BT}-1}.
\]

At resonance the integrand in Eqs. (\ref{vdwnor})  
has a large factor $\sim 1/R_i^2$,
in  sharp contrast to the case of parallel relative motion (see Eq.(\ref{vdwpar}))
, where  there is no such
enhancement factor. Thus, at resonance if $R_i^2<<1$ the friction for normal 
relative motion will be much larger than the friction for parallel relative motion.
 In  contrast to the heat transfer, the van der Waals
friction  
 is very sensitive to presence of low frequency excitations which absorb 
plenty momentum without absorbing much energy.  
 Thus the van der Waals friction is very 
sensitive to the type of the material.  

 Assuming  that the medium $\mathbf{2}$ is sufficiently rarefied and
consist of particles with the radius $r<<d$,  with the polarizability $%
\alpha(\omega)$ given by  Eq.(\ref{vdwpar}), it is easy to calculate
 friction coefficient  between a small particle and a flat surface:
\begin{eqnarray}
\Gamma_{\|} &=&\frac \hbar \pi \int_0^\infty \mathrm{d}\omega \omega \left(
-\frac{\partial n}{\partial \omega}\right) \int_0^\infty \mathrm{d}qq^2\mathrm{e}%
^{-2qd}\mathrm{Im}\alpha _2  \nonumber \\
&&\times \left\{ 2\mathrm{Im}R_{1p}\left[2+\left(\frac{\omega}
{cq}\right)^2
\right]+\left( \frac \omega {cq}\right) ^2\mathrm{Im}R_{1s}(\omega )
 \right\}  \label{fourtynine}
\end{eqnarray}
and $\Gamma_{\perp}=2\Gamma_{\|}$. In the nonretarded limit ($c\rightarrow 
\infty$) this formula agrees with the result 
obtained in \cite{Tomassone}.

For good metals ($k_{B}T/4\pi\hbar\sigma>>1$) using (\ref{vdwnor})
 for $\lambda_T(k_{B}T/4\pi\hbar\sigma)^{3/2}<d<\lambda_T(4\pi\hbar\sigma/
(k_{B}T)^{1/2}$ ($\lambda_T=c\hbar/(k_{B}T $)),
we get
\begin{equation}
\gamma_{\perp p}
\approx 0.13\frac{\hbar}{d^3\lambda_{T}}
\left(\frac{k_{B}T}{4\pi\hbar\sigma}\right)^{1/2},
\label{4fourteen}
\end{equation}
and for $d<\lambda_T(k_{B}T/4\pi\hbar\sigma)^{3/2}$ we get
\begin{equation}
\gamma_{\perp p}
\approx\frac{\hbar}{d^4}
\left(\frac{k_{B}T}{4\pi\hbar\sigma}\right)^{2}\left(1+\ln\frac{\hbar\sigma}
{2k_BT}\right).
\end{equation}
The last contribution will dominate for metal with not too high conductivity
($k_{B}T/4\pi\hbar\sigma \simeq 1$).

For  comparison, the $p$-wave contribution for parallel relative motion
 for $d<\lambda_c,\,
(\lambda_c=c/(4\pi\sigma k_{B}T)^{1/2})$
is given by \cite{Volokitin3}
\begin{equation}
\gamma_{\parallel p}^{evan}\approx 0.3\frac{\hbar}{d^4}\left(\frac{k_BT}{4\pi\hbar
\sigma}\right)^2
\label{4fifteen}
\end{equation}
It is interesting to note that, in contrast to  parallel
relative motion, for normal relative motion of good conductors,
for practically all $d>0$ the main contribution to friction comes
from retardation effects, since  Eq. (\ref{4fourteen}), in contrast to
 Eq. (\ref{4fifteen}), contains the light velocity.   

From Eq. (\ref{vdwnor}) we get the $s$-wave contribution to friction
 for $d<\lambda_c$
\begin{equation}
\gamma_{\perp s}^{evan}\approx10^{-2}\frac{\hbar}{\lambda_c^4}(3-
5\ln(2d/\lambda_c))
\label{4sixteen}
\end{equation}
For parallel relative motion the $s$-wave contribution is  two times
smaller:$\gamma_{\perp s}^{evan}=2\gamma_{\| s}^{evan}$.

Figures 5 and 6  show the calculated contribution to the friction
coefficient $\gamma $ from the evanescent electromagnetic waves
for two semi-infinite solids, with parameters chosen
to correspond to copper ($\tau ^{-1}=2.5\cdot 10^{13}s^{-1}$, $\omega
_p=1.6\cdot 10^{16}s^{-1})$ at $T=273\,K$, for parallel (Fig.5) and normal
(Fig.6) relative motion. Results are shown separately for
both the $s$- and $p$- wave contribution. The dashed line shows the result
when the local (long-wavelength) dielectric function $\epsilon (\omega
)$ is used, and the full line show the result obtained within the  non-local optic 
dielectric formalism, which was proposed some years ago for the investigation of 
the optical properties of  metals in anomalous skin effect frequency 
region \cite{Fuchs and Kliever}. 
Fig.5 shows that, for sufficiently small separations ($d<1000\,\mathrm{\AA }$%
), for parallel relative motion  the non-local optic effects become important for  the $p-$
wave contribution.  However, for the $s-$ wave contribution, for both parallel and
 normal relative motion, the non-local optic effects are
negligibly small for practically all the separations. For normal relative
motion, for the $p$-wave contribution the non-local optic effects are less important,
than for the parallel relative motion. 

For  high-resistivity metals ($k_BT/4\pi \hbar \sigma>1$) for $d<\lambda_c$ we get
\begin{equation}
\gamma_{\perp}\approx 0.48\frac{\hbar}{d^4}\frac{k_BT}{4\pi \hbar \sigma} 
\end{equation}
and $\gamma_{\|}\approx 0.1 \gamma_{perp}$. Thus, in contrast to the heat flux, the van der Waals friction diverges in the limit 
$\sigma \rightarrow 0$. Of course, in reality the friction must vanish in this 
limit since the  conductivity is proportional to concentration of free electrons, and 
the friction must vanish as the carrier concentration vanishes.
 The origin of the discrepancy 
lies in the breakdown of the macroscopic theory which was used in the calculation of 
friction at low electron concentration. The macroscopic approach for the 
electromagnetic 
properties of material is valid only when the length scale of the spatial variation of 
the electromagnetic field is much larger than the average distance 
between the electrons. 
For evanescent waves this length scale is determined by the 
separation $d$ between the bodies. Thus,  the 
macroscopic approach is valid  if $d>>n^{-1/3}$, 
where $n$ is the concentration of electrons. This fact was overlooked in Ref.\cite{
Greffet1}. From this requirement we can estimate 
the maximum friction which can be obtained for high resistivity metals. The minimum 
conductivity can be estimated as
\[
\sigma_{\min}\sim\frac{e^2\tau}{d^3m}
\]
and the maximum of friction 
\[
\gamma_{\max}\sim \frac{\hbar}{d^4}\frac{k_BT}{4\pi\hbar\sigma_{\min}} \sim 
\frac{mk_BT}{4\pi e^2\tau d}
\]

To estimate the friction coefficient $\Gamma$ for an atomic force microscope
tip with radius of curvature $r>>d$ we can use an approximate formula 
\cite{Hartmann,Apell}
\begin{equation}
\Gamma=2\pi\int_0^\infty d\rho\rho\gamma(z(\rho))
\label{approx}
\end{equation}
where it is assumed that the tip has cylinder symmetry. Here $z(\rho)$
denotes the tip - surface distance as a function of the distance $\rho$
from the tip symmetry axis, and the friction coefficient $\gamma(z(\rho))$
is determined by the expressions for  flat surfaces. 
 We assume that
the tip has a paraboloid shape given [in cylindrical coordinates ($z,\rho$)]
by the formula: $z=d+\rho^2/2r$. 
If
\begin{equation}
\gamma(\rho) = \frac{C}{\left(d+\frac{\rho^2}{2r}\right)^n}
\end{equation}
we get
\[
\Gamma=\frac{2\pi r}{n-1}\frac{C}{d^{n-1}}=\frac{2\pi rd}{n-1}\gamma(d)
=A_{eff}\gamma(d)
\]
where $A_{eff}$ is the effective surface area. For high-resistivity metals $n=4$, and 
the maximum friction coefficient for spherical tip: 
\[
\Gamma_{\max}^s \sim \gamma_{max}dr\sim
\frac{mk_BTr}{4\pi e^2\tau }
\]
Using this formula for $\tau\sim 10^{-15}s$, $r\sim 1\mu m$ and $T=300$K we 
get $\Gamma_{\max}\sim 10^{-15}kg/s$. This friction is two order of magnitude 
smaller  than  was observed in a recent experiment \cite{Stipe} at $d=10$nm. In the 
case of the cylindrical tip with the width $w$:
\[
\Gamma_{\max}^c \sim \gamma_{max}\sqrt{dr}w\sim
\frac{mk_BTwR^{0.5}}{4\pi e^2\tau d^{0.5}}
\]
For $w=7\mu$m and $d=10$nm the friction is of the same order as it was observed 
in experiment. Thus, van der Waals friction between high resistivity material can 
be measured with the present state-of-art equipment.

As in the case of the radiative heat transfer, the  van der Waals friction can be 
greatly enhanced when resonant photon tunneling  
 between localized surface 
states, e.g. surface plasmon polaritons and adsorbate vibration modes, occurs.
Using the same approximation as when deriving Eq.(\ref{seven})    for normal relative
motion we get
\begin{equation}
\gamma_{\perp}=\frac{3}{128}\frac{\hbar^2\omega_a^2}{d^4k_{B}T\eta}
\frac{1}{\sinh^2(\hbar\omega_0/2k_BT)}.
\label{5seven}
\end{equation}
Similar,  for parallel relative motion
 
\begin{equation}
\gamma_{\parallel}=\frac{\hbar^2 \eta q_c^4}{128\pi k_{B}T}
\frac{1}{\sinh^2(\hbar\omega_0/2k_BT)}
\label{5eight}
\end{equation}
where $q_c=\min(b,\ln(2\omega_a/\eta)/d)$, and where $b$ is of the order of 
an interatomic distance. Thus if
 $\ln(2\omega_a/\eta)/d>b$,  Eq. (\ref{5eight}) is independent of the 
distance.
For  small distances one can expect that $q_c$ is determined
by the dielectric properties of the material, and does not depend on $d$. In this
case the friction will be also distance independent. 

Resonant photon tunneling enhancement of the van der
Waals friction is possible for two semiconductor surfaces which can support
low-frequency surface plasmon modes. As an example we consider
two clean surfaces of silicon carbide (SiC). The
optical properties of this material were described above.
Using the same parameters as before and   at $T=300$K we get $\gamma_{\|}\approx
(10^3/d^4)$kg$\cdot$ s$^{-1}$m$^{-2}$, where the distance $d$ is in {\AA}ngstr{\o}m,
  and $\gamma_{\perp}\approx3\gamma_{\|}$.  Note that the
friction between the two semiconductors  is  about three order of magnitude
larger than between two clean  good metallic conductors (see Fig.(5,6)).

\subsection{Adsorbate vibrational mode  enhancement of the van der Waals 
friction}

Another enhancement mechanism of van der Waals friction is connected with 
resonant photon tunneling
between adsorbate vibrational modes localized on different surfaces.  
In \cite{Volokitin4,Volokitin5} we have shown that resonant photon tunneling 
between two 
surfaces covered by a low coverage of potassium atoms  at $d=1$nm gives rise to 
enhancement in 
friction by six orders of the magnitude in the comparison with friction between 
clean surfaces. 
The adsorbate induced enhancement of van der Waals friction is even larger for the 
case of Cs adsorption on Cu(100). In this case even at small coverage ($\theta\approx
0.08$) 
in adsorbed layer there is acoustic branch for vibrations parallel to the surface 
\cite{Senet}. 
 In this case at 
small frequencies the reflection coefficient is given by \cite{Volokitin} 
\begin{equation}
R_p=1-\frac{2qa\omega_q^2}{\omega^2-\omega_q^2+i\omega\eta}
\label{adsorbate1}
\end{equation}
where $\omega_q^2=4\pi n_a e^{*2}aq^2/M$, $\eta$ is the damping constant for the
 adsorbate vibrations parallel to the surface,  
$a$ is the separation between the adsorbate center and image plane. 
Using Eq.(\ref{adsorbate1}) in 
Eq.(\ref{vdwpar}) 
 for 
\[
\frac{a}{\eta d}\sqrt{\frac{4\pi n_a e^{*2}a}{Md^2}}\ll 1
\]   
we get
\begin{equation}
\gamma_{\|} \approx 0.62\frac{k_BTa^2}{\eta d^6}
\label{adsorbate2}
\end{equation}
It is interesting to note that the dependence on $n_a$, $e^*$, and $M$ is dropped out 
from Eq.(\ref{adsorbate2}). However, Eq.(\ref{adsorbate1}) is only valid when there 
 are acoustic vibrations in the adsorbed layer. For Cs adsorbed on Cu(100) surface 
the acoustic vibrations exist only for $\theta \ge 0.1$ \cite{Senet}.  The friction 
coefficient 
for the atomic force microscope can be estimated using 
approximate Eq.(\ref{approx}). Using Eq.(\ref{adsorbate2}) for a cylindrical tip we 
get 
\begin{equation}
\Gamma^c_{\|} \approx 0.68\frac{k_BTa^2r^{0.5}w}{\eta d^{5.5}}
\end{equation}
where $r$ is the radius of the curvature of the tip and $w$ is its width. In the 
case of 
Cs adsorbed on Cu(100) surface the damping constant $\eta \approx 3\cdot 10^9$s$^{-1}$ 
and $a=2.94${\AA}
\cite{Volokitin}. Than for $r=1\mu$m, $w=7\mu$m, $T=293$ K at $d=10$nm we 
get $\Gamma_{\|}=0.5\cdot 10^{-13}$kg/s, that is only three times smaller than the 
friction observed in \cite{Stipe} at the same distance. However van der Waals 
friction  is characterized by much stronger distance dependence ($\sim 1/d^{5.5}$)
 than in 
experiment ($\sim 1/d^{n}$, where $n=1.3\pm 0.2$). Thus at smaller distances the 
van der Waals friction will be much larger than friction observed in \cite{Stipe} 
and can be measured experimentally. Fig.7 shows  
the friction coefficient between the 
copper tip and the copper substrate as a function of the separation $d$, 
when the surfaces of the tip and the substrate 
are covered by low concentration of the Cs atoms and for the clean surfaces. In 
comparison, the friction between two clean surfaces at the separation $d=1$nm is 
eleven 
orders of the magnitude smaller. However, the friction between clean surfaces 
shown in Fig.(7) was calculated in local optic approximation. For parallel relative 
motion the non-local 
optic effects are very important  (see Fig.5) and when these non-local optic 
effects are taken into account  
the  friction between  adsorbate covered surfaces at $d=1$nm will be 
by seven orders of magnitude larger than the friction between clean surfaces 
in parallel 
relative motion.

\section{Summary}

We have studied how the radiative heat transfer and van der Waals friction 
between two bodies 
depends on the dielectric
properties of the media. We have found that, at  short distances
between the bodies, the thermal flux 
can be significantly enhanced in comparison
with the black body radiation, in particular when the material involved can support
low-frequency adsorbate vibrational modes, or surface plasmon modes, or the
conductivity of the metals is chosen to optimize the heat transfer. This
fact can be used in the scanning probe microscopy for local heating and
modification of surfaces.
We have shown that the van der Waals friction can be enhanced by several 
orders of magnitude in the case of resonant photon tunneling between low-
frequency surface plasmon modes and adsorbate vibrational modes. 
In the case of friction between two Cu(100) surfaces covered by a low concentration 
of cesium  atoms at $d=10$nm we have found the friction to be of the same order 
of the magnitude as it was observed in experiment \cite{Stipe}. 
However, the van der Waals 
friction is characterized by stronger distance dependence than the friction observed in 
experiment. Thus at small distances the van der Waals friction can be much larger 
than the friction observed in \cite{Stipe} and can be measured experimentally. This 
effect can be important technologically  for ultrasensitive force registration and 
from basic point of view. The friction observed in \cite{Stipe} can be explained 
by \textit{electrostatic} friction \cite{Volokitin} when the electromagnetic 
field in the vacuum gap is mediated by bias voltage or by inhomogeneities of the 
surfaces.

\vskip 0.5cm \textbf{Acknowledgment }

A.I.V acknowledges financial support from DFG and Russian Foundation for 
Basic Research (Grant N 04-02-17606) B.N.J.P. acknowledges
support from the European Union Smart Quasicrystals project.

\vskip 0.5cm

FIGURE CAPTIONS

Fig. 1 (a) The heat transfer flux between two semi-infinite silver bodies as
a function of the separation $d$, one at temperature $T_1=273\ \mathrm{K}$
and another at $T_2=0\ \mathrm{K}$. (b) The same as (a) except that we have
reduced the Drude electron relaxation time $\tau $ for solid 1 from a value
corresponding to a mean free path $v_F\tau =l=560\ \mathrm{\AA }$ to $20\
\mathrm{\AA }$. (c) The same as (a) except that we have reduced $l$ to $3.4\
\mathrm{\AA }$. For silver (Fig.1a) at $T=273$ K the conductivity $\sigma=
5.6\cdot 10^{17}$s$^{-1}$ and $k_BT/4\pi \hbar \sigma =4.6\cdot 10^{-6}$ and for 
Fig.1b and Fig.2c these quantities can be obtained using scaling $\sigma 
\sim l^{-1}$. (The base of the logarithm is 10)

Fig. 2 The thermal flux as a function of the conductivity of the solids. The
solid surfaces are separated by $d=10\ \mathrm{\AA }$. The heat flux 
 for other separations can be obtained using scaling
$\sim 1/d^2$ which holds for high-resistivity materials. (The base of the
logarithm is 10)

Fig. 3. The heat flux between two semi-infinite silver bodies coated with $%
10\,\mathrm{\AA }$ high resistivity ($\rho =0.14\ \mathrm{\Omega cm}$)
material. Also shown is the heat flux between two silver bodies, and two
high-resistivity bodies. One body is at zero temperature and the other at $%
T=273K$. (a) and (b) shows the $p$ and $s$-contributions, respectively. (The
base of the logarithm is 10)

Fig. 4. The heat flux between two surfaces covered by adsorbates 
and  between two clean surface, as a
function of the separation $d$. One body is at zero temperature and the
other at $T=273\,$K. For parameters corresponding to K/Cu(001) and 
Cu(001) \cite{Senet} (%
$\omega_{\perp}=1.9\cdot 10^{13}s^{-1}, \omega_{\parallel}= 4.5\cdot
10^{12}s^{-1}, \eta_{\parallel}=2.8\cdot 10^{10}s^{-1},
\eta_{\perp}=1.6\cdot 10^{12}s^{-1}, e^{*}=0.88e$) (The
base of the logarithm is 10)

Fig. 5. The friction coefficient for two flat surfaces in parallel relative
motion as a function of
separation $d$ at $T=273\,$K with parameter chosen to correspond to copper ($%
\tau ^{-1}=2.5\cdot 10^{13}s^{-1},\,\omega _p=1.6\cdot 10^{16}s^{-1}$). The
contributions from the $s-$ and $p-$polarized electromagnetic field are
shown separately. The full curves represent the results obtained within the
non-local optic dielectric formalism, and the dashed curves represent the
result obtained within local optic approximation. (The
base of the logarithm is 10)

Fig. 6. The friction coefficient for two flat surfaces in normal relative
motion as a function of
separation $d$ at $T=273\,$K with parameter chosen to correspond to copper ($%
\tau ^{-1}=2.5\cdot 10^{13}s^{-1},\,\omega _p=1.6\cdot 10^{16}s^{-1}$). The
contributions from the $s-$ and $p-$polarized electromagnetic field are
shown separately. The full curves represent the results obtained within the
non-local optic dielectric formalism, and the dashed curves represent the
result obtained within local optic approximation. (The
base of the logarithm is 10)

Fig. 7. The friction coefficient between the copper tip and copper substrate which 
surfaces are covered by low concentration of cesium atoms, as a
function of the separation $d$.  The cylindrical tip is characterized by radius 
of curvature $r=1\mu$m and width $w=7\mu$m.  For other parameters corresponding 
to Cs adsorbed on Cu(100) surface at coverage $\theta \approx 0.1$
 and for Cu(100)  \cite{Senet,Volokitin}: $e^*=0.28e$, $\eta=3\cdot 10^{9}$s$^{-1}$, 
$a=2.94${\AA}, $T=293$ K. (The
base of the logarithm is 10)

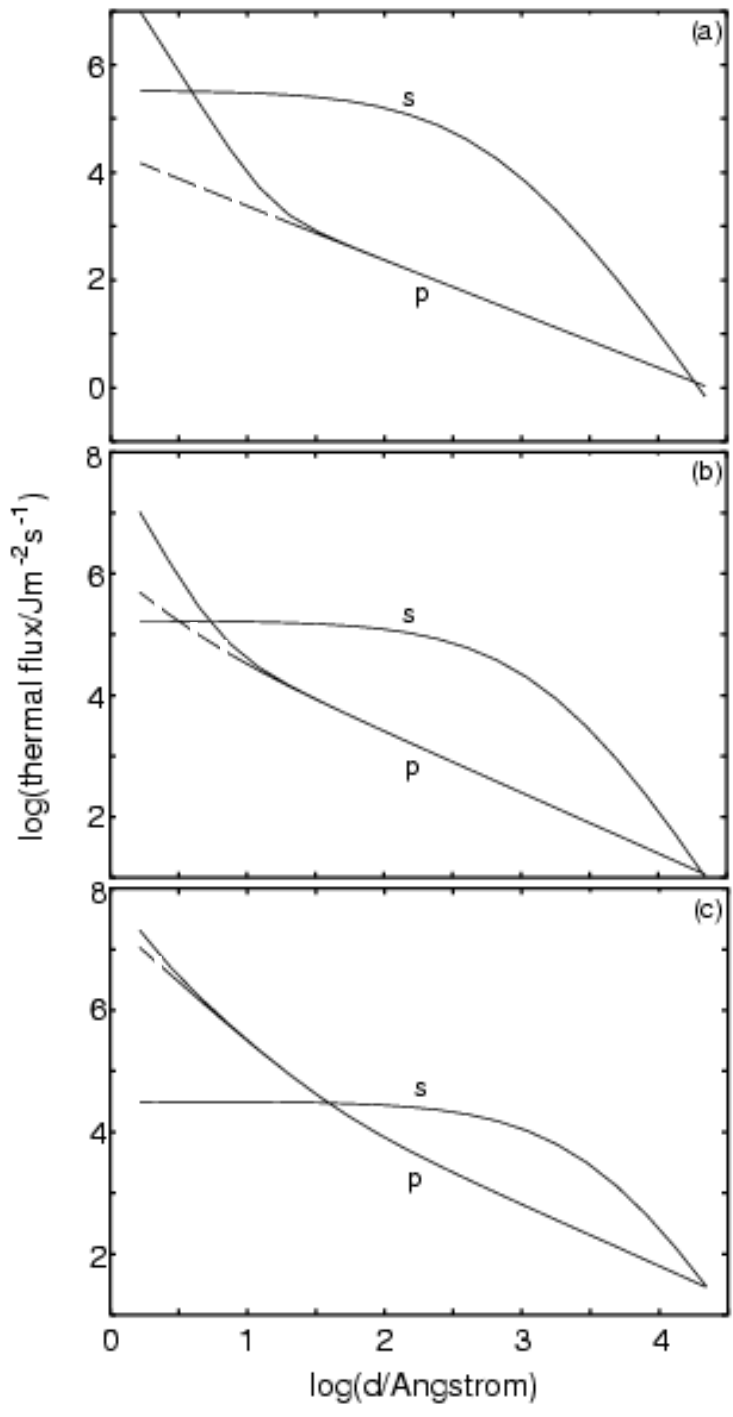
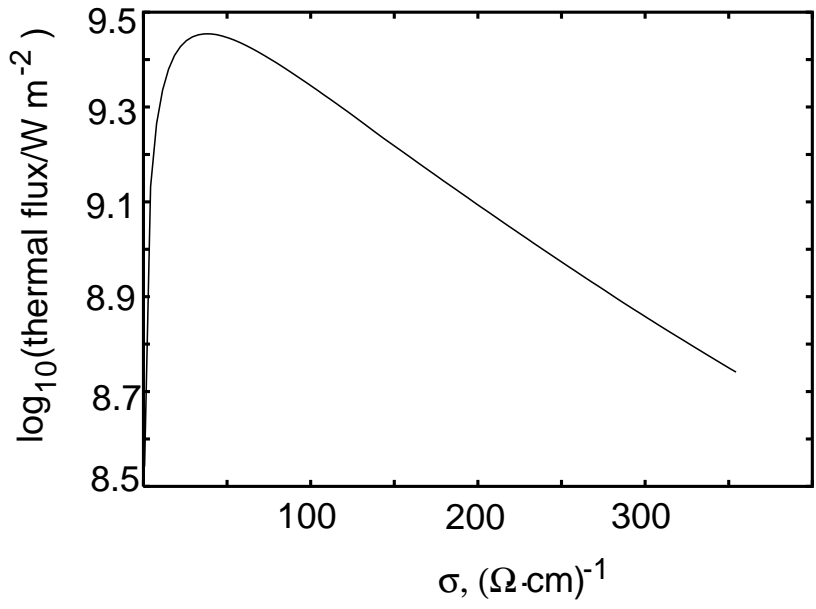
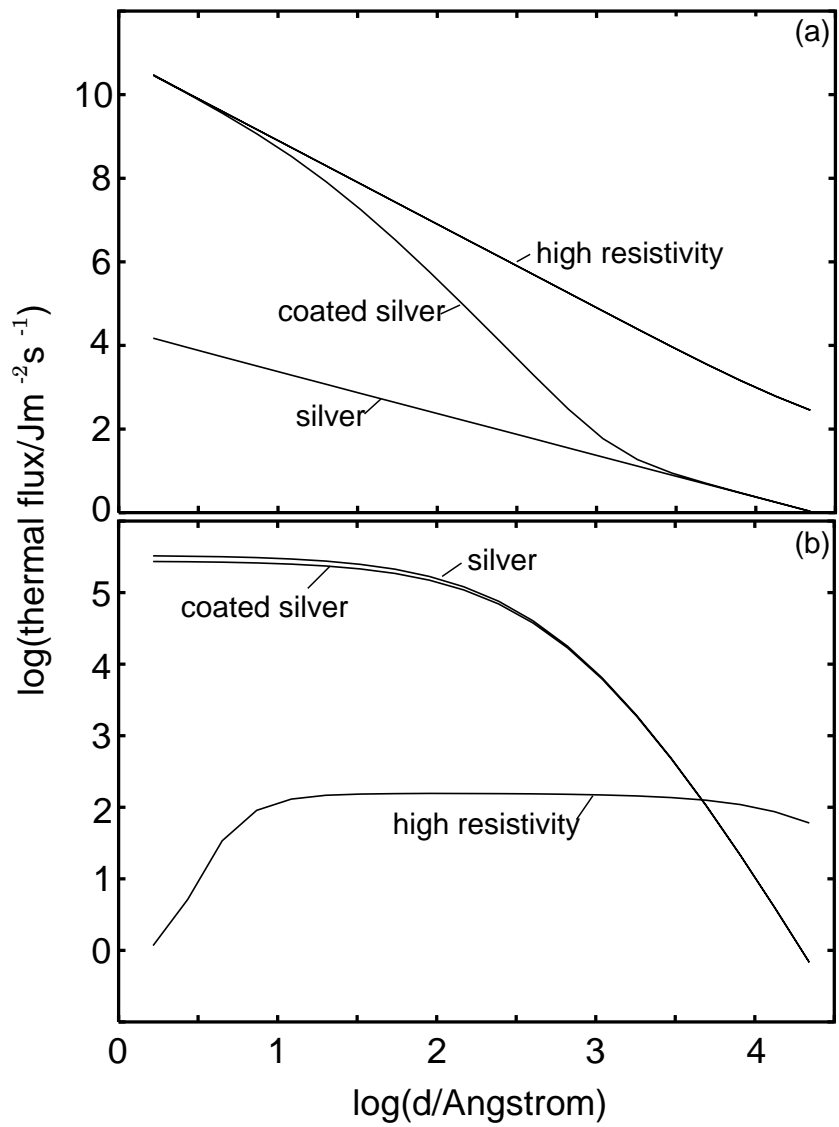
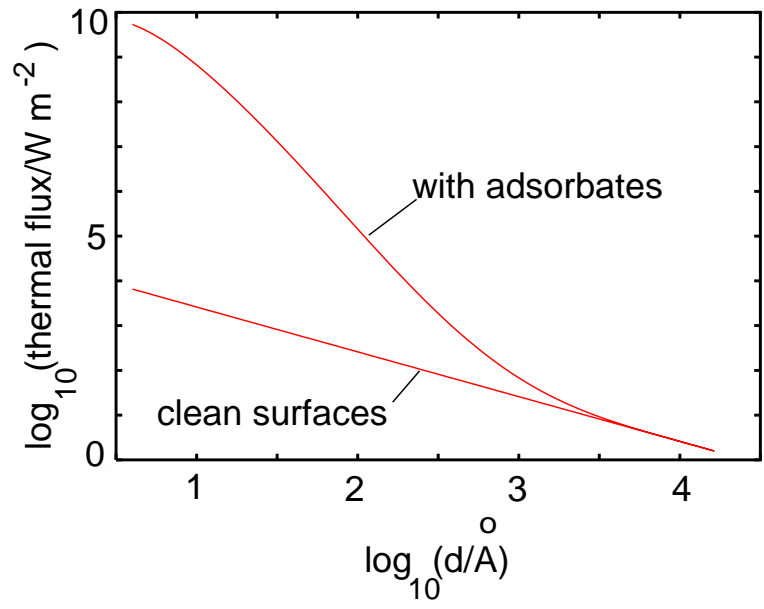
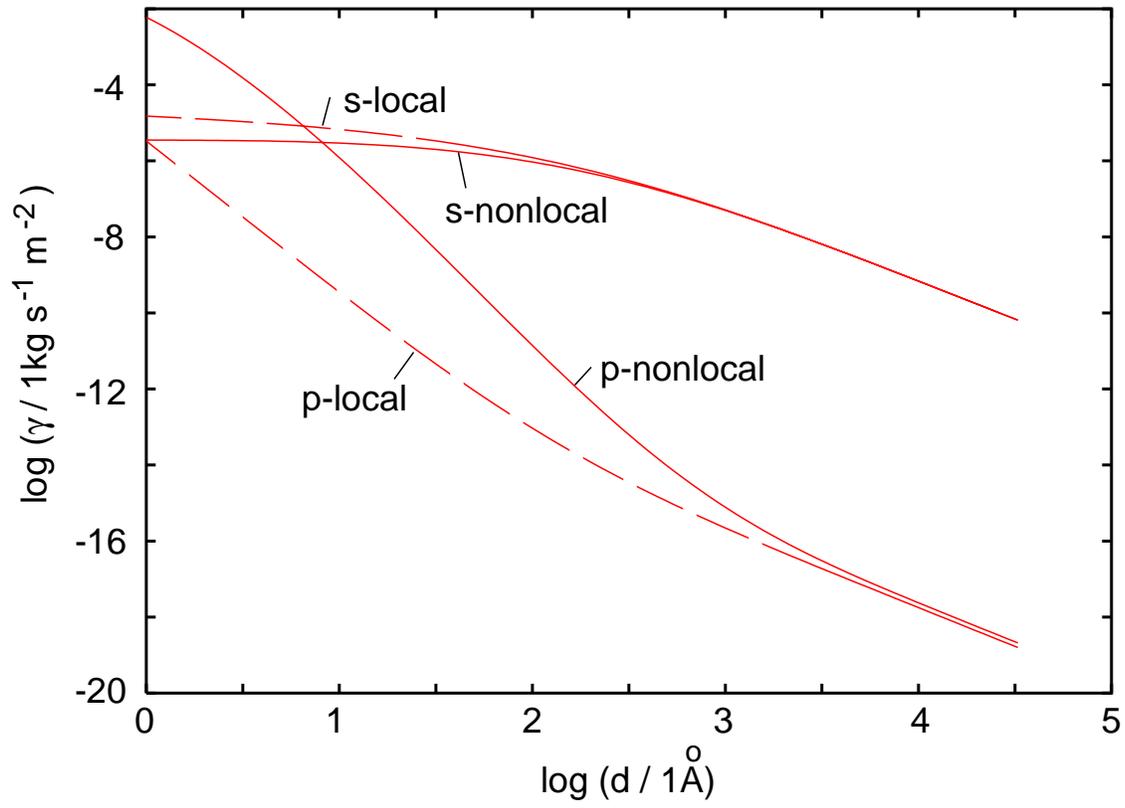
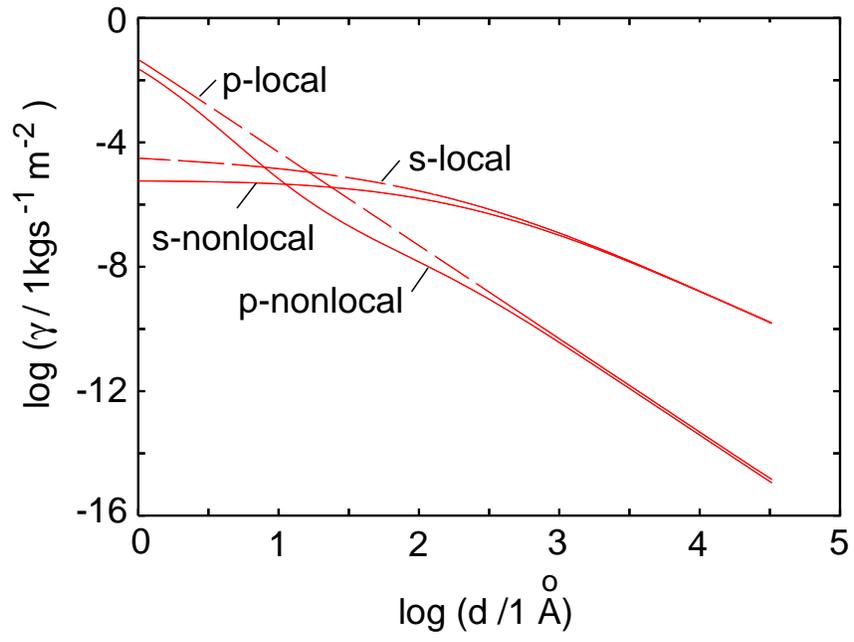
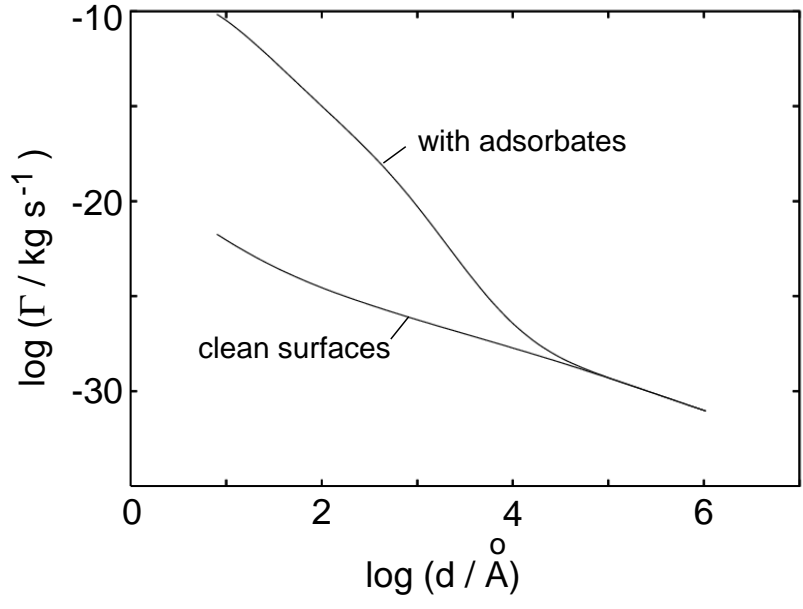
\end{document}